\documentclass[twocolumn]{emulateapj}
\usepackage{graphicx}
\usepackage{dcolumn}

\begin{document}

\title{A {\it Chandra} Snapshot Survey of IR-bright LINERs:  A Possible Link Between Star Formation, AGN Fueling, and Mass Accretion}

\author{R. P. Dudik, \altaffilmark{1} S. Satyapal, \altaffilmark{1,2}  M. Gliozzi, \altaffilmark{1} \& R. M. Sambruna\altaffilmark{1,3}}

\altaffiltext{1}{George Mason University, Department of Physics \& Astronomy, MS 3F3, 4400 University Drive, Fairfax, VA 22030}

\altaffiltext{2}{Presidential Career Award Scientist}

\altaffiltext{3}{George Mason University, School of Computational Sciences, MS 5C3, 4400 University Drive, Fairfax, VA 22030}

\begin{abstract}
We present results from a high resolution X-ray imaging study of nearby LINERs observed by ACIS on board {\it Chandra}.  This study complements and extends previous X-ray studies of LINERs, focusing on the under-explored population of nearby dust-enshrouded infrared-bright LINERs.  The sample consists of 15 IR-bright LINERs (L$_{\rm FIR}$/L$_{\rm B}$ $>$ 3), with distances that range from 11 to 26 Mpc.  Combining our sample with previous {\it Chandra} studies we find that $\sim$ 51\% (28/55) of the LINERs display compact hard X-ray cores. The nuclear 2-10 keV luminosities of the galaxies in this expanded sample range from $\sim$ 2 $\times$ 10$^{38}$ ergs s$^{-1}$ to $\sim$ 2 $\times$ 10$^{44}$ ergs s$^{-1}$.  We find that the most extreme IR-faint LINERs are exclusively AGN.  The fraction of LINERs containing AGN appears to decrease with IR-brightness and increase again at the highest values of L$_{\rm FIR}$/L$_{\rm B}$.  We find that of the 24  LINERs showing compact nuclear hard X-ray cores in the expanded sample which were observed at H$\alpha$ wavelengths, only 8 actually show evidence of a broad line.  Similarly, of the 14 LINERs showing compact nuclear hard X-ray cores with corresponding radio observations, only 8 display a compact flat spectrum radio core.  These findings emphasize the need for high-resolution X-ray imaging observations in the study of IR-bright LINERs.  Finally, we find an intriguing trend in the Eddington ratio vs. L$_{\rm FIR}$ and L$_{\rm FIR}$/L$_{\rm B}$ for the AGN-LINERs in the expanded sample that extends over seven orders of magnitude in L/L$_{\rm Edd}$.  This correlation may imply a link between black hole growth, as measured by the Eddington ratio, and the star formation rate (SFR), as measured by the far-IR luminosity and IR-brightness ratio.  If the far-IR luminosity is an indicator of the molecular gas content in our sample of LINERs, our results may further indicate that the mass accretion rate scales with the host galaxy's fuel supply.  We discuss the potential implications of our results in the framework of black hole growth and AGN fueling in low luminosity AGN. 

\end{abstract}

\section{Introduction}
One third to as many as one half of all galaxies in the local Universe are classified as Low Ionization Nuclear Emission Line Regions (LINERs), displaying narrow optical emission lines of low ionization uncharacteristic of photoionization from normal stars (e.g., Heckman 1980; Ho, Filippenko, \& Sargent 1997a; Stauffer 1982).  Since the low excitation optical line spectrum can be produced by a range of processes including shock heating (e.g. Dopita et al. 1996) through cloud-cloud collisions in galaxy mergers or surrounding active galactic nuclei (AGN), cooling flows, and photoionization by nonthermal energy sources (e.g. Ferland \& Netzer 1983) or hot stars (e.g. Terlevich \& Melnick 1985), the nature of the central engine in LINERs has been under debate for the past several decades. Because LINERs dominate the population of extragalactic objects in the nearby Universe, determining the nature of their central engine has considerable bearing on a number of issues of astrophysical interest.  Indeed, since normal galaxies are now known ubiquitously to host quiescent black holes (Gebhart et al. 2000; Ferrarese \& Merritt 2000), the low mass accretion rates inferred for many confirmed accretion-powered LINERs may suggest that these objects capture the population of AGN just before accretion onto the black hole ``turns off".  As a consequence, establishing the number of LINERs containing accreting black holes (``AGN-LINERs"), their luminosities, accretion rates, and the relationship of these quantities with the properties of the parent galaxy will provide critical insight into some of the most fundamental questions in extragalactic research today.

The vast majority of LINER galaxies are dust-enshrouded, emitting the bulk of their radiation at infrared (IR) wavelengths.  Using the compilation by Carillo et al. (1999) of all known LINERs studied in the literature, approximately 80\% of all nearby LINERs have far-IR luminosities exceeding that in the B-band\footnote[1]{Far-infrared luminosities (in units of solar luminosities: L$\odot$) correspond to the 40-500$\mu $m wavelength interval and were calculated using the IRAS 60 and 100 $\mu $m fluxes according to the prescription: L$_{\rm FIR}$=1.26$\times$10$^{-14}$(2.58f$_{60}$+f$_{100}$) in W m$^{-2}$ (Sanders \& Mirabel 1996), L$_{B}$: B magnitude see Carrillo et al (1999)}. However, almost all detailed multiwavelength investigations of nearby LINERs have been based on optically-selected samples, which generally exclude the IR-bright population (e.g. radio: Nagar et al. 2002, X-ray: Terashima et al. 1998, 2003, Ho et al. 2001; optical: Ho et al.1997b).  While considerable progress has been made on understanding these sources, very little is known about the more numerous IR-bright population. Any study of the AGN detection rate and properties of accretion powered LINERs based on optically-selected samples will therefore be biased toward the IR-faint sources.  Furthermore, IR-bright LINERs, characterized by dusty host galaxies with higher star formation rates, may possibly capture a distinct subset or evolutionary state of the LINER population.   By expanding the sample of LINERs studied to include a larger range of IR-brightness ratios, an understanding of possible evolutionary sequences as well as the physical significance of IR-brightness in the LINER class can for the first time be explored.

To investigate the properties of the IR-bright population, we undertook a {\it Chandra} ``snapshot" survey of a sample of nearby IR-bright LINERs.  Traditional optical or ultraviolet studies of IR-bright LINERs can be inconclusive, since highly obscured AGN in dusty galaxies can be hidden at these wavelengths. Sensitive hard X-ray observations at high spatial resolution with {\it Chandra} can greatly aid in determining accurately the number of LINERs containing accreting black holes, providing a robust probe of obscured AGN out to column densities of a few times 10$^{24}$ cm$^{-2}$.  Detection of a single compact hard X-ray source coincident with the nucleus in the most nearby galaxies would be the telltale sign of an accreting central black hole.  In this paper, we build on our previous archival {\it Chandra} study of LINERs (Satyapal, Sambruna, \& Dudik 2004; henceforth SSD04) as well as previous {\it Chandra} studies of LINERs (e.g. Ho et al. 2001, henceforth H01; Eracleous et al. 2002;) in order to increase the sample of IR-bright LINERs and obtain a more comprehensive understanding of this important class of objects.

The outline of this paper is the following.  In Section 2, we summarize the sample selection criteria for the new observations presented in this work and describe the basic properties of the LINER sample used in our analysis.  In Section 3, we present our observations and data reduction techniques, together with a tabulation of X-ray luminosities and morphological descriptions.  In Section 4, we discuss the analysis and assumptions adopted in deriving black hole masses, bolometric luminosities, and Eddington ratios. In Section 5 we present our results, including a discussion of our derived AGN detection rates, a comparison with other AGN indicators, X-ray morphological findings, luminosities, and correlations between various quantities.  In Section 6 we discuss the implications of our correlations and summarize our conclusions in Section 7.

\section{The Sample}

The survey sample was selected from the multi-frequency LINER catalog from Carillo et al. (1999), which consists of an extensive database of 476 LINERs compiled from the literature. We selected all nearby IR-bright sources that satisfy the following criteria:  1) L$_{\rm FIR}$/L$_{\rm B}$ $>$ 3 and 2) D $<$ 30 Mpc (H$_0$ = 75 km s$^{-1}$ Mpc$^{-1}$, q$_0$ = 0.5), excluding the few objects that meet these criteria that were already observed by {\it Chandra}.  Our targets range in distance from approximately 8 to 26 Mpc, with a median distance of 14 Mpc.  The sample, consisting of a total of 16 objects, is summarized in Table 1a.  We note that one IR-faint galaxy, NGC 4350, was accidently included in our program.  Fourteen of our targets have been optically classified as LINERs according to the Veilleux \& Osterbrock (1987) diagnostic diagrams.   NGC 3125 and IC 1218 are classified as LINERs according to the Heckman (1980) criteria.

In order to enlarge the statistics of our present analysis, and explore the physical significance of IR-brightness in the LINER class, we combined our sample with the LINER/transition objects from SSD04 and H01.  The entire {\it Chandra} dataset contains 58 LINERs.  In this paper, we refer to this dataset as the ``expanded sample.''  We emphasize that this expanded sample is heterogenous and not complete and therefore subject to selection biases.  Also, our definition of an IR-bright LINER as one with L$_{\rm FIR}$/L$_{\rm B}$ $>$ 3 is largely arbitrary; the LINERs in the expanded sample form a continuous distribution in L$_{\rm FIR}$/L$_{\rm B}$ that ranges from 0.1 to 162.9.  The basic properties of the expanded sample are summarized in Figure 1 and Tables 1a, 1b, \& 1c.  We conducted a literature search to determine which objects were observed at optical and radio wavelengths, listing in Table 1 all LINERs displaying broad H$\alpha$ emission lines or flat spectrum radio cores suggestive of an AGN.  

\section{Observations and Data Reduction Procedure}

{\it Chandra} observations of our targets were obtained with the Advanced CCD Imaging Spectrometer (ACIS-S) with the source at the nominal aim point of the S3 CCD.  The exposures were 5ks per target. In some cases, background flares were detected and had to be subtracted from the original 5ks exposure, resulting in shorter exposure times for these objects. All of the observations were carried out using the standard 3.2s frame time.  For all but one observation, the nuclear counts were insufficient for photon pileup to be significant.   The observation of NGC 7465 experienced severe pileup which affected the countrate by 45\% or more and the data were thus determined to be unreliable.  This galaxy has been excluded from our results and discussion.\\

The {\it Chandra} data were processed using CIAO v.2.3 using the latest calibration files provided by the {\it Chandra} X-ray Center (CXC).  In Table 2 we list the details of the {\it Chandra} observations. Our data reduction procedure and analysis follow the treatment described by H01.  The 0.3-8keV energy range was chosen for analysis.  Nuclear count rates were extracted from a 2" radius centered on the nucleus, where the position was determined by radio observations, if available, or 2MASS observations.  The coordinates used for the extraction are listed in Table 2.  The background was extracted from a nearby circular region of radius 30" free of spurious X-ray sources.  Detections are defined when the X-ray counts are at least 3$\sigma$.  The upper limits listed in Tables 3a, correspond to 3$\sigma$ values corresponding to a 2" extraction centered on the radio or infrared nucleus, with the background estimated as discussed above for the case of the detections. \\

For three of the sixteen sources in our proprietary sample, NGC 3125, NGC 4102, and NGC 5005, the detected counts in the 0.3-8 keV range were sufficient to employ detailed spectral fits.  Spectral fitting was performed using XSPEC v.11.2.0 (Arnaud 1996).  We began by fitting all three sources to a single power-law model with the absorption column density fixed at the Galactic value.  This base model was then modified with additional components until the minimum {\it $\chi$}$^{2}$ was obtained.  The errors on the best-fit parameters are at 90\% confidence ($\Delta${\it $\chi$}$^2$ = 2.7) for each parameter of interest.  We discuss these models in Section 5.4.  In the case of objects for which detected counts in the 0.3-8 keV regime were insufficient for spectral fitting, the nuclear count rate was converted to 2-10 keV X-ray luminosities assuming a canonical intrinsic power-law spectrum with photon index $\Gamma$ = 1.8 using the Galactic interstellar absorption listed in Table 1\footnote[2]{We adopt the same power-law slope as H01 which is typical for low luminosity AGN ($\Gamma$ ranges from 1.6-2.0; Terashima \& Wilson 2003; see also Section 5.4 of this work)}. We also explored the morphological characteristics of our X-ray sample following the classification scheme adopted by H01.  These results are discussed in Section 5.3.

\begin{figure*}[]
{\includegraphics[width=8cm]{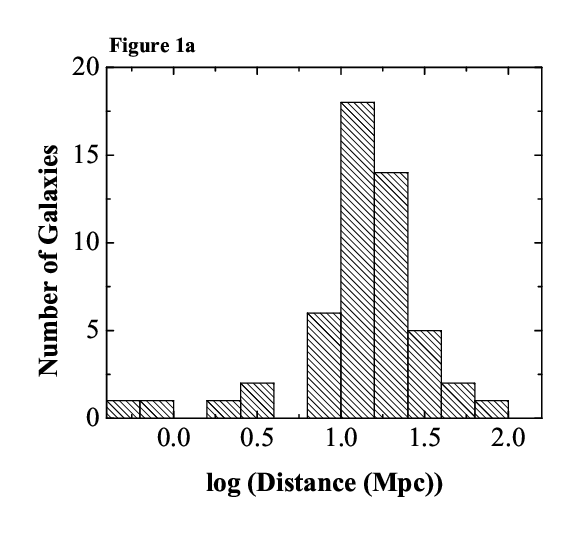}}{\includegraphics[width=8cm]{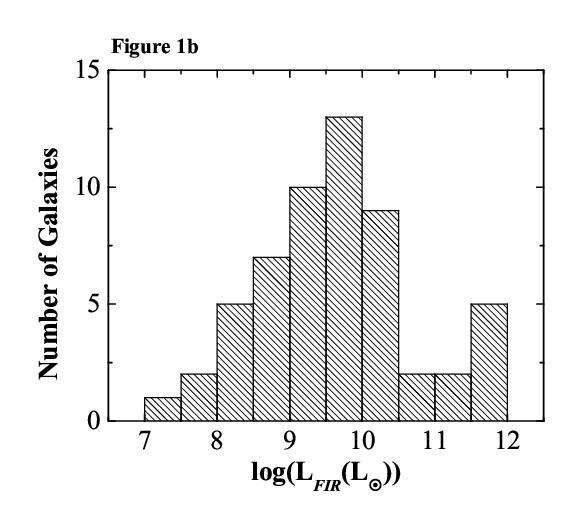}}\\
{\includegraphics[width=8cm]{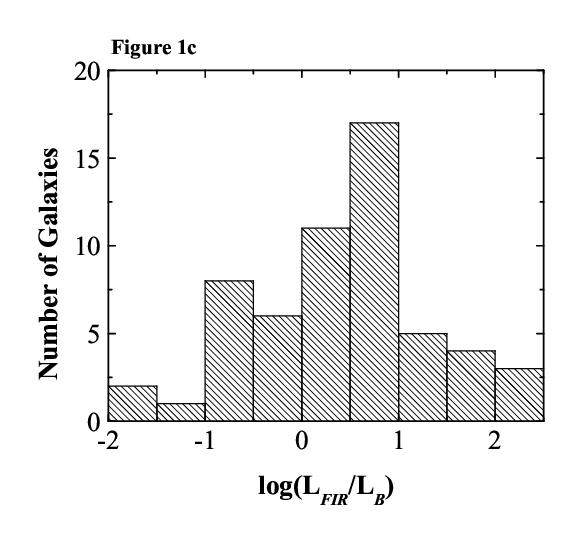}}{\includegraphics[width=8cm]{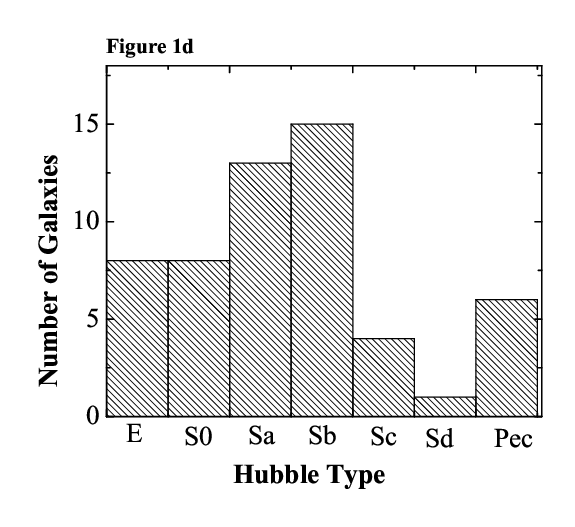}}\\
\caption[]{Characteristics of the expanded {\it Chandra} sample of LINERs from this paper, SSD04, \& H01.  Most galaxies are nearby and span a wide range of luminosities, IR-brightness ratios, and Hubble types.}
\end{figure*}

\begin{table*}
\begin{center}
\begin{tabular}{lccccccccc}
\multicolumn{10}{l}{\bf Table 1: Properties of the Expanded Sample}  \\ 
 \hline\hline
\multicolumn{10}{l}{\bf {\it 1a: New {\it Chandra} Observations: Sample Properties}}  \\ 
 \hline
\multicolumn{1}{c}{Galaxy} & \multicolumn{1}{c}{Distance} & 
Hubble & Optical & log & L$_{\rm FIR}$/ & Broad & FRS & 
{\it N}$_{\rm H}$ & log(M$_{\rm BH}$) \\ 
\multicolumn{1}{c}{Name} & \multicolumn{1}{c}{(Mpc)} & Type & 
Class & (L$_{\rm FIR}$) & L$_{\rm B}$ & H$_{\alpha}$$?$ &  & cm$^{-2}$
& (M{\scriptsize $\odot $}) \\ 
\multicolumn{1}{c}{(1)} & \multicolumn{1}{c}{(2)} & (3) & (4) & (5) & (6) & 
(7) & (8) & (9) & (10) \\ 
 \hline\hline
NGC 0660 & 11.3 & SB(s)a;p. & T & 10.1 & 34.4 & no$^a$ & no$^d$ & 4.9 & 7.35\\
NGC 1055 & 13.3 & SBb;sp & T & 9.9 & 7.7 & no$^a$ & $\cdots$ & 3.4 & 6.53\\
NGC 3125 & 11.5 & S & L* & 9.0 & 9.0 & $\cdots$ & $\cdots$ & 5.7 & 5.77\\
NGC 4013 & 11.1 & Sb & T & 9.2 & 5.2 & no$^a$ & $\cdots$ & 1.4 & $\cdots$\\
NGC 4102 & 11.2 & SAB(s)b? & T & 10.0 & 23.5 & no$^a$ & $\cdots$ & 1.8 & $\cdots$\\
NGC 4350$^{\dagger}$ & 16.6 & SA0 & L & 8.3 & 0.2 & no$^a$ & no$^d$ & 2.6 & 7.99\\
NGC 4419 & 16.8 & SB(s)a & T & 9.6 & 5.0 & no$^a$ & no$^d$ & 2.7 & 6.94\\
NGC 4527 & 23.2 & SAB(s)bc & T & 10.4 & 8.7 & no$^a$ & no$^d$ & 1.9 & 8.25\\
NGC 4666 & 20.3 & SABc & T & 10.4 & 11.7 & no$^b$ & no$^e$ & 1.7 & $\cdots$\\
NGC 4713 & 8.7 & SAB(rs)d & T & 8.8 & 3.2 & no$^a$ & no$^d$ & 2.0 & $\cdots$\\
NGC 4750 & 21.6 & (R)SA(rs)ab & L & 9.7 & 3.4 & yes$^3$ & $\cdots$ & 1.7 & $\cdots$\\
NGC 5005 & 12.6 & SAB(rs)bc & L & 9.8 & 3.6 & yes$^3$ & no$^d$ & 1.1 & $\cdots$\\
NGC 5678 & 25.6 & SAB(rs)b & T & 10.1 & 6.5 & no$^a$ & $\cdots$ & 1.3 & $\cdots$\\
NGC 5954 & 26.2 & SAa:;p. & T & 10.1 & 18.5 & $\cdots$ & $\cdots$ & 3.3 & 6.81\\
IC 1218 & 14.8 & S? & L* & 8.5 & 4.8 & $\cdots$ & $\cdots$ & 4.1 & $\cdots$\\
NGC 7465$^{\star}$  & 26.3 & (R')SB(s)0: & T & 9.7 & 7.2 & $\cdots$ & $\cdots$ & 6.0 & 7.59$^+$\\
\hline\hline
\end{tabular}
\end{center}
\end{table*}
\begin{table*}
\begin{center}
\begin{tabular}{lccccccccc}
\multicolumn{10}{l}{\bf Table 1: Properties of the Expanded Sample cont.}  \\ 
 \hline\hline
\multicolumn{10}{l}{\bf {\it 1b: SSDO4: Sample Properties}}  \\ 
 \hline
\multicolumn{1}{c}{Galaxy} & \multicolumn{1}{c}{Distance} & 
Hubble & Optical & log & L$_{\rm FIR}$/ & Broad & FRS & 
{\it N}$_{\rm H}$ & log(M$_{\rm BH}$) \\ 
\multicolumn{1}{c}{Name} & \multicolumn{1}{c}{(Mpc)} & Type & 
Class & (L$_{\rm FIR}$) & L$_{\rm B}$ & H$_{\alpha} $$?$ &  & cm$^{-2}$ & (M{\scriptsize $\odot $}) \\ 
\multicolumn{1}{c}{(1)} & \multicolumn{1}{c}{(2)} & (3) & (4) & (5) & (6) & 
(7) & (8) & (9) & (10) \\ 
 \hline\hline
0248+4302 & 205.2 & Gpair & L & 11.5 & 138.3 & no$^a$ & $\cdots$ & 1.0 & $\cdots$\\
1346+2650 & 253.0 & cD;S0? & L & 10.6 & 3.5 & no$^a$ & $\cdots$ & 1.2 & 9.07$^{+}$\\
2312-5919 & 178.4 & Merger & $\cdots$ & 11.7 & 89.6 & no$^b$ & $\cdots$ & 2.8 & $\cdots$\\
2055-4250 & 171.3 & Merger & $\cdots$ & 11.7 & 66.6 & no$^b$ & $\cdots$ & 3.9 & $\cdots$\\
IC1459 & 22.6 & E3 & L & 8.6 & 0.1 & $\cdots$ & $\cdots$ & 1.2 & 9.00$^g$\\
MRK266NE & 112.2 & Compact;p.& L & 11.2 & 25.8 & no$^a$ & $\cdots$ & 1.7 & $\cdots$\\
MRK273 & 151.1 & Ring gal. & L & 11.8 & 162.9 & no$^b$ & $\cdots$ & 1.1 & 7.74\\NGC0224 & 0.8 & SA(s)b & $\cdots$ & 9.0 & 0.5 & $\cdots$ & $\cdots$ & 0.1 & 7.57\\
NGC0253 & 2.6 & SAB(s)c & $\cdots$ & 9.9 & 9.8 & $\cdots$ & $\cdots$ & 1.4 & 7.07\\
NGC0404 & 2.4 & SA(s)0 & L & 7.3 & 0.6 & no$^a$ & no$^f$ & 5.3 & 5.88\\
NGC0835 & 54.3 & SAB(r)ab;p. & L & 10.6 & 9.0 & $\cdots$ & $\cdots$ & 2.2 & 8.97\\
NGC1052 & 29.6 & E4 & L & 9.1 & 0.3 & yes$^c$ & $\cdots$ & 3.1 & 8.29$^g$\\
NGC3031 & 3.6 & SA(s)ab & L & 8.4 & 0.1 & yes$^c$ & yes$^d$ & 4.2 & 7.79$^g$\\
NGC3079 & 15.0 & SB(s)c & L* & 10.3 & 16.6 & no$^a$ & $\cdots$ & 7.9 & 7.65$^g$\\
NGC3368 & 12.0 & SAB(rs)ab & L & 9.4 & 1.0 & no$^a$ & no$^d$ & 2.8 & 7.16\\
NGC3623 & 10.8 & SAB(rs)a & L* & 9.0 & 0.5 & no$^a$ & no$^d$ & 2.5 & 8.16\\
NGC4125 & 18.1 & E6;pec & T & 8.6 & 0.1 & no$^a$ & $\cdots$ & 1.8 & 8.50\\
NGC4278 & 16.1 & E1-2 & L & 8.5 & 0.2 & yes$^c$ & yes$^d$ & 1.8 & 9.20$^g$\\
NGC4314 & 12.8 & SB(rs)a & L & 9.0 & 1.2 & no$^a$ & no$^d$ & 1.8 & 7.22\\
NGC4374 & 18.4 & E1 & L & 8.5 & 0.3 & no$^a$ & yes$^d$ & 2.6 & 9.20$^g$\\
NGC4486 & 16.0 & E+0-1;p. & L & 8.3 & 0.0 & no$^a$ & yes$^d$ & 2.5 & 9.48$^g$\\
NGC4569 & 16.8 & SAB(rs)ab & T & 9.7 & 1.1 & no$^a$ & no$^f$ & 2.5 & 7.58\\
NGC4579 & 16.8 & SAB(rs)b & L & 9.5 & 0.9 & yes$^c$ & yes$^f$ & 2.5 & 7.85$^g$\\NGC4696 & 39.5 & E+1;p. & L & $\cdots$ & $\cdots$ & $\cdots$ & no$^d$ & 8.1 & 8.60\\
NGC5194 & 8.4 & SA(s)bc;p. & L & 9.8 & 1.7 & no$^a$ & no$^d$ & 1.6 & 6.90$^g$\\
NGC5195 & 7.7 & SB01;p. & L* & 9.4 & 3.0 & no$^a$ & no$^d$ & 1.6 & 7.90\\
NGC6240 & 97.9 & I0:;p. & L & 11.3 & 38.2 & $\cdots$ & $\cdots$ & 5.8 & 9.15\\
NGC6500 & 40.1 & SAab & L & 9.4 & 1.4 & no$^a$ & yes$^f$ & 7.4 & 8.82$^g$\\
NGC6503 & 0.6 & SA(s)cd & T & 6.8 & 2.1 & no$^a$ & no$^d$ & 4.1 & 5.53\\
NGC7331 & 11.0 & SA(s)b & T & 9.8 & 3.8 & no$^a$ & no$^d$ & 8.6 & 7.91\\
UGC05101 & 157.6 & S? & L & 11.7 & 118.5 & no$^a$ & $\cdots$ & 4.2 & $\cdots$\\
\hline\hline
\end{tabular}
\end{center}
\end{table*}
\begin{table*}
\begin{center}
\begin{tabular}{lccccccccc}
\multicolumn{10}{l}{\bf Table 1: Properties of the Expanded Sample cont.}  \\ 
 \hline\hline
\multicolumn{10}{l}{\bf {\it 1c: H01 Sample Properties}}  \\ 
 \hline
\multicolumn{1}{c}{Galaxy} & \multicolumn{1}{c}{Distance} & 
Hubble & Optical & log & L$_{\rm FIR}$/ & Broad & FRS & 
{\it N}$_{\rm H}$ & log(M$_{\rm BH}$) \\ 
\multicolumn{1}{c}{Name} & \multicolumn{1}{c}{(Mpc)} & Type & 
Class. & (L$_{\rm FIR}$) & L$_{\rm B}$ & H$_{\alpha} $$?$ &  & cm$^{-2}$ & (M{\scriptsize $\odot $}) \\ 
\multicolumn{1}{c}{(1)} & \multicolumn{1}{c}{(2)} & (3) & (4) & (5) & (6) & 
(7) & (8) & (9) & (10) \\ 
 \hline\hline
NGC 2787 & 7.5 & SB0 & L & 7.8 & 0.3 & yes$^c$ & yes$^d$ & 4.26 & 7.59$^g$\\
NGC 2841 & 12.0 & Sb: & L & 9.2 & 0.5 & no$^a$ & yes$^d$ & 1.48 & 8.42$^g$\\
NGC 3489 & 12.1 & SAB0 & T & 9.2 & 1.1 & no$^a$ & no$^d$ & 1.96 & 7.58\\
NGC 3627 & 10.3 & SABb & T & 10.1 & 4.1 & no$^a$ & yes$^d$ & 2.29 & 7.26$^g$\\
NGC 3628 & 10.3 & Sb p. & T & 10.0 & 6.1 & no$^a$ & no$^d$ & 2.25 & 7.91$^g$\\
NGC 3675 & 12.8 & Sb & T & 9.6 & 3.1 & no$^a$ & no$^d$ & 2.13 & 7.11$^g$\\
NGC 4203 & 15.0 & SAB0: & L & 8.5 & 0.4 & yes$^c$ & yes$^d$ & 1.18 & 7.90$^g$\\
NGC 4321 & 16.1 & SABbc & T & 10.1 & 2.8 & no$^a$ & no$^d$ & 2.32 & 6.81$^g$\\
NGC 4494 & 17.1 & E1 & L & 7.7 & 0.0 & no$^a$ & no$^d$ & 1.54 & 7.65$^g$\\
NGC 4594 & 9.8 & Sa & L & 9.0 & 0.2 & no$^a$ & $\cdots$ & 2.80 & 9.04$^g$\\
NGC 4826 & 7.5 & Sab & T & 9.6 & 1.8 & no$^a$ & no$^d$ & 2.63 & 7.74\\
& & & & & & & & & \\ 
\hline\hline
\end{tabular}
\end{center}

{\noindent{\scriptsize {\bf Columns Explanation:} Col(1):Common Source Names; $^{\star}$ NGC 7465 experienced severe pileup and has been excluded from our results and discussion; $\dagger$ This galaxy is IR-faint and was accidently observed in our program.   Col(2): Distance (for H$_{0}$= 75 km s$^{-1}$Mpc$^{-1}
$; the distance of NGC 4419 was taken from Ho et al. 1997b); Col(3): Morphological Class; Col(4): Optical classification scheme:  All galaxies in this sample were classified using the Veilleux et al. 1995 / Ho et al. 1997 optical classification scheme: (LINERs, L = [OI]$\lambda$6300 $\geq$ 0.17 H$\alpha$, [NII]$\lambda$6583 $\geq$ 0.6 H$\alpha$, [SII]$\lambda\lambda$6716,6731 $\geq$ 0.4 H$\alpha$, [OIII]$\lambda$5007/H$\beta$ $<$ 3 (Veilleux 1995); Transition objects, T, obey all of the former ratios with the additional criteria 0.08 H$\alpha$ $\leq$ [OI]$\lambda$6300 $<$ 0.17 H$\alpha$ (Ho 1997);  With the exception of NGC 3125 and IC 1218 which were classified using the Heckman 1980 scheme: L$^*$ = [OII]$\lambda$3727 $<$ [OIII]$\lambda$5700 and [OI]$\lambda$6300 $<$ 0.33 [OIII]$\lambda$5700.)  Col(5): Far-infrared luminosities (in L$\odot $) correspond to the 40-500$\mu $m wavelength interval calculated using the prescription of Sanders \& Mirabel 1996; Col(6): L$_{B}$: B magnitude see Carrillo et al (1999); Col(7): LINERs with broad H$\alpha $ emission; Col(8): LINERs with a flat radio spectrum (NGC 4419 shows a steep radio spectrum.) \ Col(9): Galactic {\it N}$_{\rm H}$(in units of$\times $ 10$^{20}$ cm$^{-2}$); Col(10): Mass of central black hole calculated using the stellar velocity despersion in the formula:  M$_{\rm BH}$ =  1.2($\pm$0.2)$\times$10$^8$ M$_{\odot}$($\sigma$$_e$/200 km s$^{-1}$)$^{3.75 ({\pm} 0.3)}$)  (from Ferrarese \& Merritt 2000; Gebhardt et al. 2000; Tremaine et al. 2002), $^+$Calculated using the prescription of Kormendy \& Richstone 1995; Ferrarese \& Merritt 2000; \& Gebhardt et al. 2003, log(M$_{\rm BH}$) = -1.58($\pm$2.09) - 0.488($\pm$0.102)M$_{B}$.}}

{\noindent{\scriptsize {\bf References: }$^{a}$ Ho, Filipenko, \& Sargent
1997a; $^{b}$ Veilleux et al 1995; $^{c}$ Ho et al. 1997b; $^{d}$
Nagar et al. 2002; $^{e}$ Dahlem et al. 1997; $^{f}$ Nagar et al. 2000; $^{g}$ Merloni, Heinz, \& Di Matteo 2003 and references therein.}}

\end{table*}

\section{Analysis}

In order to gain a better physical understanding of nuclear activity in LINERs and its relation to IR-brightness, we explored the relationship between IR-brightness and the fundamental quantities of black hole mass, bolometric luminosity and mass accretion rate for all confirmed AGN-LINERs in our expanded sample.  The assumptions adopted and associated uncertainties are described below.

\subsection{Black Hole Masses}
Published black hole masses exist and were adopted in this work for 19 of the 58 LINERs listed in Table 1.  These estimates were obtained for the most part through direct modeling of stellar and gas kinematics. For those galaxies with measured central velocity dispersions, we estimated the black hole masses using the tight correlation between black hole mass and stellar velocity dispersion (Ferrarese \& Merritt 2000; Gebhardt et al. 2000; Tremaine et al. 2002), demonstrated to hold for nearby AGN (Ferrarese et al. 2001; McLure \& Dunlop 2002):
\begin{equation}
M_{\rm BH} =  1.2(\pm0.2)\times{10^8} M_{\odot}(\sigma_e/200 {\rm km s^{-1}})^{3.75 (\pm0.3).}
\end{equation}
Here, $\sigma$$_e$ is the luminosity-weighted line-of-sight velocity dispersion within the half-light radius and is within 10\% of the central velocity dispersion.  Central velocity dispersion measurements were found for 24 additional objects.  All measurements were obtained from the catalog of central velocity dispersions listed in the Hypercat database\footnote[3]{Velocity dispersions taken from the Hypercat database available online at http://www-obs.univ-lyon1.fr/hypercat} (Paturel et al. 1997). For all LINERs without central velocity dispersion measurements in elliptical hosts, we used the correlation between the bulge magnitude and the black hole mass (Kormendy \& Richstone 1995; Ferrarese \& Merritt 2000; Gebhardt et al. 2003):
\begin{equation}
log(M_{\rm BH}) = -1.58(\pm2.09) - 0.488(\pm0.102)M_{\rm B},
\end{equation}
where M$_{BH}$ is in solar masses and M$_{\rm B}$ is the total B-band magnitude of the host elliptical galaxy.  Since the bulge magnitude is highly uncertain in spirals, we only employed equation 2 for the 2 ellipticals in our sample without central velocity dispersion measurements (NGC 7465 \& MRK273, SSD04).

\subsection{Bolometric Luminosities}

In order to estimate the bolometric luminosities of our targets from our X-ray observations, knowledge of the bolometric correction factor appropriate for the LINER AGN class is required.  While the broad-band Spectral Energy Distribution (SED) of traditional higher luminosity AGN are relatively well-studied and follow a fairly universal shape (Elvis et al. 1994), the SEDs of the nuclear source in AGN-LINERs is not as well-known.  In the few cases where broad spectral coverage of the nuclear source is available, the SEDs of LINERs are found to differ markedly from those of conventional AGN (Ho 1999), resembling instead the SEDs predicted for radiatively inefficient accretion flows (e.g. Quataert \& Narayan 1999; Narayan et al. 2002).  From the SED measurements of seven low luminosity AGN from Ho (1999), the 2-10 keV luminosity ranges from 2 to 9 \% of the integrated bolometric luminosity.  Using the mean value of L$_{\rm bol}$=34 $\times$ L$_{\rm X}$(2-10keV), we estimated the bolometric luminosities of all of our AGN targets using the nuclear 2-10 keV luminosities listed in Table 3.  We note that these estimates are somewhat uncertain and a determination of the true error in this factor requires more extensive studies of the SEDs of the LINER class.

Once the black hole mass has been estimated, the Eddington luminosity can be calculated using L$_{\rm Edd}$ = 1.26 $\times$10$^{38}$M$_{\rm BH}$/M$_{\odot}$ ergs s$^{-1}$ (e.g., Rees 1984; Peterson 1997).  The bolometric luminosity of the AGN is directly proportional to the accretion rate, L$_{\rm bol}$ =  $\epsilon$$\dot{M}$$_{\rm acc}$c$^2$, and therefore the ratio, L/L$_{\rm Edd}$ $\propto$  $\dot{M}$$_{\rm acc}$/$\dot{M}$$_{\rm Edd}$, is an indirect measure of the accretion rate relative to the critical Eddington value.

\section{Results}
\subsection{AGN Detection Rate}
Twenty-eight LINERs from the expanded LINER sample display a hard compact nuclear X-ray point source coincident with the {\it VLA} or {\it 2MASS} nucleus.  The minimum 2-10 keV luminosity of those objects displaying hard nuclear cores in the entire sample is 2 $\times$ 10$^{38}$ ergs s$^{-1}$ (NGC 2787; H01).  Young supernova remnants, X-ray-binaries (XRBs), and/or hot diffuse gas from starburst-driven winds are known to emit X-rays, however these sources are weak and/or are usually spatially extended at the distances of our objects.  For example, while intergalactic XRBs are known to reach luminosities of $\sim$ 10$^{38}$ ergs s$^{-1}$ (McClintock \& Remillard 2003),  typical XRBs have luminosities up to 10$^{37}$ ergs s$^{-1}$ (White, Nagase, \& Parmar 1995).   Several tens to over ten thousand XRBs, concentrated in a spatial extent of only a few tens of parsecs, would be required to produce the luminosities of the hard nuclear point sources, which is highly implausible.  Moreover, the detection of a single hard nuclear point source coincident with the {\it VLA} or {\it 2MASS} nucleus is highly suggestive of an AGN.  Therefore, in this paper, we define AGN-LINERs as all LINERs displaying hard nuclear cores with luminosites L$_{\rm X}$ (2-10keV) $\geq$ 2 $\times$ 10$^{38}$ ergs s$^{-1}$.   

\begin{figure*}[]
{\includegraphics[width=8.5cm]{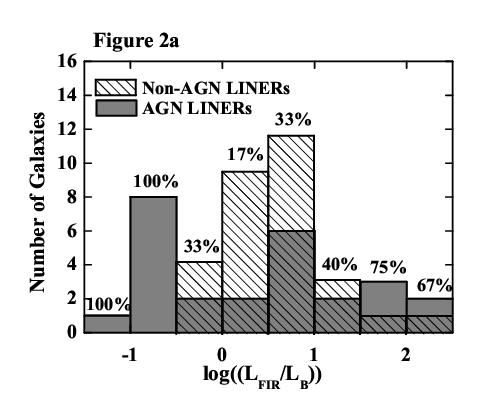}}{\includegraphics[width=8.5cm]{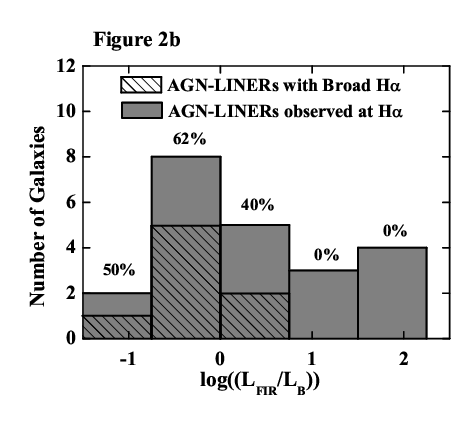}}\\
\caption[]{2a: X-ray derived detection rate as a function of IR-brightness.  Above each column we mark the percentage of observed galaxies displaying AGN with respect to the total number of galaxies in a given luminosity ratio bin.  As can be seen, IR-faint LINERs are exclusively AGN.  The fraction of LINERs containing AGN appears to decrease with IR-brightness and then increase again at the most extreme L$_{\rm FIR}$/L$_{\rm B}$ values.  2b: Fraction of LINERs displaying H$\alpha$ as a function of IR-brightness.  Again above each column we mark the percentage of observed galaxies displaying a broad H$\alpha$ line with respect to the total number of galaxies in a given luminosity ratio bin.  As can be seen, most of the LINERs that show the broad H$\alpha$ line are IR-faint and objects with higher L$_{\rm FIR}$/L$_{\rm B}$ values show no evidence of AGN activity at optical wavelengths, emphasizing the importance of high spatial resolution X-ray observations in the study of IR-bright LINERs.}
\end{figure*}

Our IR-bright LINER survey complements the SSD04 and H01 surveys with respect to L$_{\rm FIR}$/L$_{\rm B}$ values and allows us to examine the X-ray-derived ``AGN detection rate'' as a function of IR-brightness. In Figure 2a we show the X-ray ``AGN detection rate'' as a function of IR-brightness.  Our results indicate, with limited statistics,  that the most extreme IR-faint LINERs are exclusively AGN.  The fraction of LINERs containing AGN appears to decrease with IR-brightness and increase again at the most extreme L$_{\rm FIR}$/L$_{\rm B}$ values.   This result may simply indicate that most IR-faint galaxies, being generally deficient in dust and gas, are bulge-dominated systems, where there are not many sources of excitation other than an AGN that can produce a LINER-like spectrum.  IR-bright galaxies, however, are generally dusty disk-dominated systems, where the presence of hot stars and starburst-driven shocks can easily give rise to a LINER spectrum (e.g., Isobe \& Feigelson, 1992).  LINERs with the most extreme L$_{\rm FIR}$/L$_{\rm B}$ values are generally ultraluminous galaxies (e.g., Sanders \& Mirabel 1996), where the large concentration of dust and gas in these often advanced merging systems conceivably increases the likelihood that an AGN is responsible for the LINER spectrum. 

\subsection{Comparison with other AGN Indicators}

Detection of broad H$\alpha$ emission can be a powerful AGN diagnostic at optical wavelengths.  However, broad optical lines can be ambiguous AGN indicators since they are dependent on viewing angle and dust obscuration, which is particularly a problem in IR-bright and low luminosity sources (Nagar et al. 2000).  Indeed, broad lines have also been produced in starburst models (Terlevich et al. 1995).  We searched the literature for optical observations at H$\alpha$ wavelengths and list those galaxies which show broad H$\alpha$ emission in Table 1.  We compare the H$\alpha$ and X-ray AGN diagnostics in Figure 2b.  There are 24 X-ray classified AGN-LINERs observed at the H$\alpha$ wavelength.  Only 33\% (8/24) show the broad lines, most of which are IR-faint. The objects with higher L$_{\rm FIR}$/L$_{\rm B}$ values show no evidence of AGN activity at optical wavelengths, emphasizing the importance of high spatial resolution X-ray observations in the study of IR-bright LINERs.  

Compact flat spectrum radio cores can also be a signature for accretion onto a black hole.  We searched the literature for radio observations of the objects in our expanded sample.  As can be seen from Table 1, 33 LINERs have corresponding radio observations.  Ten of these sources show flat spectrum radio cores.  Interestingly, two of these objects, NGC 2841 and NGC 3627, show no sign of AGN activity in the X-ray.  In these cases, thermal emission from optically thin ionized gas in compact nuclear starbursts (Condon et al. 1991) can give rise to the observed radio properties from these sources.  Detection of a flat spectrum radio core does not necessarily discriminate between nuclear starbursts and accretion onto black holes unless the brightness temperature exceeds 10$^5$ K (Condon et al. 1991).  In the case of NGC 2841 and NGC 3627, the brightness temperature limit of T$_b$ $>$ 10$^{3}$-10$^{4}$K and T$_b$ $>$ 10$^{2.5}$-10$^{3}$K, respectively, is consistent with either a starburst nucleus or an AGN (Nagar et al. 2000).  

Of the 28 LINERs with AGN signatures in the X-ray, only 14 have corresponding radio observations.  Roughly half of these (8/14) show a flat spectrum radio core. A nuclear flat radio spectrum is not always a clear indicator of an AGN.  For instance, flat spectrum compact radio cores are found in only $\sim$10\% of ``classical" Seyfert galaxies (de Bruyn \& Wilson 1978; Ulvestad \& Wilson 1989; Sadler et al. 1995).  In addition, while flat spectrum radio cores are known to be present in many elliptical galaxies (Heckman 1980), they are uncommon in spirals (Villa et al. 1990; Sadler et al. 1995).  This limits the diagnostic capability of radio observations in our heterogenous LINER sample, again emphasizing the importance of high spatial resolution X-ray observations.

\subsection{X-ray Morphologies}
  The X-ray images of all of the LINERs in this study have been classified into four morphological types according to the scheme adopted by H01.  These classes are defined as follows:  class (I) objects exhibit a dominant hard nuclear point source, class (II) objects exhibit multiple hard off-nuclear point sources of comparable brightness to the nuclear source, class (III) objects reveal a hard nuclear point source embedded in soft diffuse emission, and class (IV) objects display no nuclear source.  Morphological class designations for all galaxies in our expanded sample are listed in Tables 3a, 3b, \& 3c.  We note that we adopt these morphological class designations for comparative purposes.  These designations are somewhat subjective and furthermore depend on the detection limits for the three samples included in the analysis in this paper.   We identify class (I) and class (III) objects as AGN-LINERs.  Class (II) objects have off-nuclear sources of comparable brightness to the nuclear source. Therefore, these objects are more likely to be morphologically consistent with starburst galaxies, expected to contain a large population of young stars and scattered XRBs.  

Seven of our objects (class IV) show no nuclear X-ray source.  The flux limit for our exposures is 1 $\times$ 10$^{-15}$ ergs s$^{-1}$ cm$^{-2}$.  This limit corresponds to a luminosity of  L$_{\rm X}$ (2-10keV)  $\approx$ 1 $\times$ 10$^{37}$ ergs s$^{-1}$ assuming an intrinsic power law spectrum of photon index 1.8 and a median Galactic absorption of 2 $\times$ 10$^{20}$ cm$^{-2}$ for the median sample distance of $\sim$ 14 Mpc.  This luminosity limit is at least an order of magnitude fainter than the weakest hard nuclear sources detected by H01 or SSD04.

Our non-detections imply either: 1) the lack of an energetically significant AGN, or 2) a highly obscured AGN.  Assuming these objects harbor an AGN at least as luminous as our lowest luminosity AGN-LINER (2 $\times$ 10$^{38}$ ergs s$^{-1}$), we can calculate the column density necessary to obscure an AGN in our non-detections (class IV objects.) Using our adopted power law of $\Gamma$ = 1.8, the required column densities were calculated and are listed in Table 3.   These column densities range from $\sim$ 3 $\times$ 10$^{22}$ to 1 $\times$ 10$^{24}$ cm$^{-2}$.   We note that the majority of AGN-LINERs are more luminous than 2 $\times$ 10$^{38}$ ergs s$^{-1}$.  The average luminosity of an AGN-LINER in the expanded sample is $\approx$ 9 $\times$ 10$^{42}$ ergs s$^{-1}$.  The column density required to obscure an AGN of this luminosity is also listed in Table 3.  As can be seen the column densities are extremely high - in all cases in excess of 10$^{24}$ cm$^{-2}$.  All but two of the upper limits in the expanded sample are below our lowest luminosity AGN-LINER.  The upper limits on these two class (IV) objects (NGC3623, SSD04 \& NGC4321 H01), do not allow us to classify them as either AGN or non-AGN LINERs.  These galaxies are therefore excluded from our calculations of the AGN detection rate and from Figures 2 and 3.  

\begin{figure*}[]
\begin{center}
{\includegraphics[width=9.5cm]{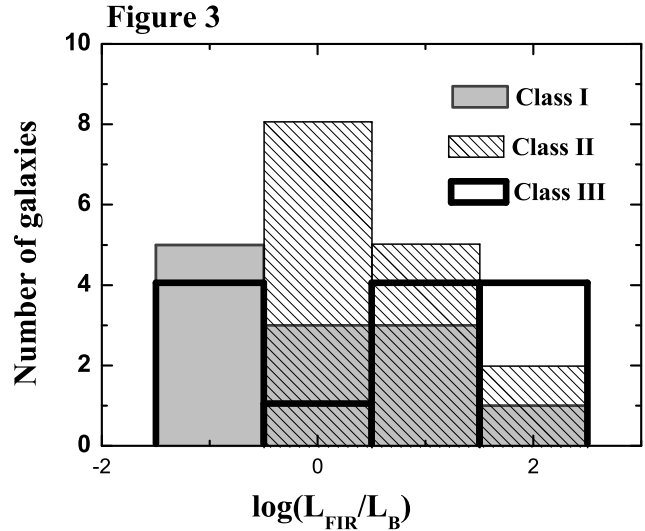}}\\
\caption[]{Morphological class as a function of IR-brightness.}
\end{center}
\end{figure*} 
In Figure 3 we compare our X-ray morphological designations with the IR-brightness of the objects.  Plotted in this histogram, are the AGN-LINERs (class (I) and class (III) objects) and the detected non-AGN-LINERs (Class (II) objects).  We find:
\begin{enumerate}
\item Those objects with the lowest IR-brightness ratio are always AGN-LINERs (class (I) and (III)). \item Class (I) AGN-LINERs become less abundant as the IR-brightness ratio increases. \item The most extreme IR-bright objects are mostly class (III) objects.  We note that this point is a {\it suggestive trend}.  At present, few galaxies occupy the extreme IR-brightness bins. More data at either extreme of IR-brightness is therefore necessary to confirm the trend.   This may indicate that class (III) objects, in addition to an AGN, contain circumnuclear starbursts, which would be expected for exceedingly IR-bright objects. \item Class (II) objects, which have nuclei typical of starburst galaxies (non-AGN-LINERs) occupy and dominate the intermediate L$_{\rm FIR}$/L$_{\rm B}$ regime.  If these galaxies are starbursts they would not be expected to be extremely IR-faint. 
\end{enumerate}

\subsection{X-ray Luminosities}

\begin{figure*}[]
\begin{center}
{\includegraphics[width=9cm]{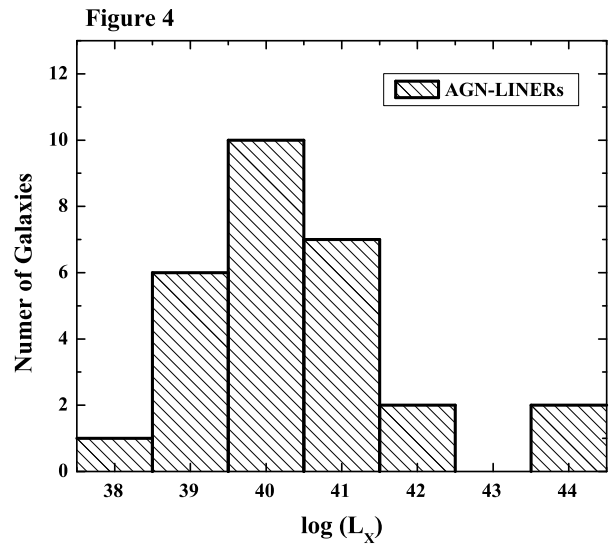}}\\
\end{center}
\caption[]{Range of X-ray luminosities for the expanded sample.}
\end{figure*}
 
From Tables 3a, 3b, \& 3c, the X-ray luminosities (2--10keV) of the AGN-LINERs in our expanded sample range from $\sim$ 2 $\times$ 10$^{38}$ to $\sim$ 2 $\times$ 10$^{44}$ ergs s$^{-1}$.  The full range of the X-ray luminosities in the combined samples is plotted in Figure 4.  We note that the majority of objects occupy the 10$^{39}$ to 10$^{41}$ ergs s$^{-1}$ luminosity range, higher than our luminosity threshold of L$_X$ (2-10keV) = 2 $\times$ 10$^{38}$, and well above all of the upper limits in all three samples.

The majority of the 2-10 keV X-ray luminosities listed in Tables 3a, 3b, \& 3c, were calculated assuming a power law with photon index $\Gamma$ = 1.8 using the Galactic absorption listed in Table 1.   Three of the fifteen LINERs (NGC 3125, NGC 4102, \& NGC 5005) had counts sufficient to allow spectral fits, allowing us to assess the accuracy of the generic power-law model.  In these models, the photon index ranged from 1.5 to 2.0 with an average of 1.8.  We note that our average value and the value for $\Gamma$ used in our generic power law model coincides with the value adopted by H01 and the average value found in low luminosity AGN (Terashima et al. 2002).  This range (1.5-2.0) in photon index has a marginal impact on the derived 2-10 keV flux, corresponding to a factor of less than 1.5 difference in our calculated luminosities. The models for the three galaxies are given in detail below.

{\it NGC 3125}:  This galaxy's spectrum was fit using a single power law model with $\Gamma$ = 2.0$^{+0.4}$$_{-0.3}$. The absorption column density was fixed at the Galactic value and an additional intrinsic absorption component was included in the model .   The resulting  intrinsic absorption for the best fit model was {\it N}$_{\rm H}$ = (5 $\pm$ 2.5) $\times$ 10$^{21}$ cm$^{-2}$, which is significant at greater than 99\% confidence level.  The additional intrinsic absorption component corresponds to a factor of  less than 2 difference in the luminosity calculated using our generic model ($\Gamma$ = 1.8, Galactic absorption).  The reduced {\it $\chi$}$^{2}$  in our best fit model is 1.0 (13 Degrees of Freedom, d.o.f.).

{\it NGC 5005}:   This galaxy was initially fit with  a single power-law model ($\Gamma$ = 1.9$^{+0.4}$$_{-0.2}$) with absorption column density fixed at the Galactic value.  The resulting poor fit ({\it $\chi$}$^{2}$$_{\rm reduced}$ = 2.44 with 7 d.o.f.) and the presence of a clear excess at soft energies, indicated the need for an additional component.   A thermal component was applied (kT = 0.90$^{+0.2}$$_{-0.3}$ keV, {\it Z}/{\it Z}$_{\odot}$ = 1.0 fixed), which yielded an acceptable fit for this galaxy ({\it $\chi$}$^{2}$$_{\rm reduced}$ = 0.94 with 9 d.o.f.).  The thermal component for NGC 5005 affects the flux by less than 3\%.

{\it NGC 4102}:  This galaxy too was initially fit with  a single power-law model ($\Gamma$ = 1.5$^{+0.6}$$_{-0.4}$) with absorption column density fixed at the Galactic value.  The resulting poor fit ({\it $\chi$}$^{2}$$_{\rm reduced}$ = 3.9 with 9 d.o.f.), and again the presence of a clear excess at soft energies, indicated the need for an additional component.  A thermal component was applied (kT = 0.54 $\pm$ 0.4 keV, {\it Z}/{\it Z}$_{\odot}$ = 0.20).  This fit was not adequate, however for NGC 4102.  It was further improved ($\Delta${\it $\chi$}$^2$ of 12.5 for 2 additional degrees of freedom) by adding a Gaussian component at a fixed energy of 6.4 keV in the source rest frame.  The line is significant at a 90\% confidence level, according to an F-test (but please see Protassov et al. 2002 for a discussion of using this method to assess line significance.)  However, the energy range for the data set is not sufficient to determine the line parameters accurately.  The residuals above 5-6keV show a clear excess, which cannot be ascribed to the background.  This excess can be fit adequately with a Gaussian component, which does not necessarily imply the presence of a physical iron line at 6.4keV.  Deeper {\it Chandra} observations or observations using the spectral capabilites of {\it XMM} would be needed to accurately assess the physical significance of this component.  The thermal component for NGC 4102 affects the flux by less than 1\%.  The reduced {\it $\chi$}$^{2}$ for this galaxy is 1.5 (5 d.o.f.).  NGC 4102 is difficult to compare with the other objects in our sample because of the presence of the Gaussian component.

Assuming a zeroth order approximation that all AGN-LINERs have roughly the same intrinsic spectrum, the results obtained from the spectral analysis of NGC 3125 can be used to infer additional spectral information for the objects with lower countrates for which only a hardness ratio analysis could be performed\footnote[4]{The hardness ratios are defined here as (H-S)/(H+S), where H represents the hard counts and S represents the soft counts in the nucleus}  Since NGC4350, NGC 4419, and NGC 4527 have hardness ratios similar to NGC 3125, which showed the presence of a local absorption component, we hypothesize that those sources also have an absorption component in excess of the Galactic value.  Similarly, NGC 660 and NGC 4750 have hardness ratios comparable to NGC 5005, which showed a clear excess at soft energies and required a thermal component in addition to the generic model.  NGC 660 and NGC 4750 are likely to contain a similar thermal component in addition to Galactic absorption in their model.  Thus our generic power-law model is likely to be a reasonably accurate model for all of the galaxies in our sample.

In Figures 5a and 5b we plot the X-ray flux versus the far-IR flux and the IR-brightness ratio.  Plotting fluxes has the advantage of avoiding spurious correlations introduced by distance effects.  As can be seen, we find no correlation in either plot.  We applied a Spearman rank test to each of these plots (Kendall \& Stuart 1976).  A Spearman rank coefficient of 1 or -1 indicates a strong correlation and a value of zero indicates no correlation.  For Figure 5a, the Spearman rank coefficient is r$_S$=-7 $\times$ 10$^{-3}$.  For Figure 5b, we find a Spearman rank coefficient of r$_S$=-0.24.

\begin{figure*}[]
{\includegraphics[width=8.5cm]{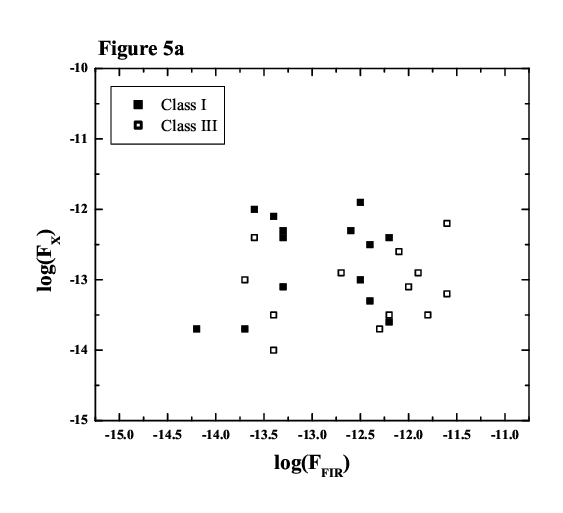}}{\includegraphics[width=8.5cm]{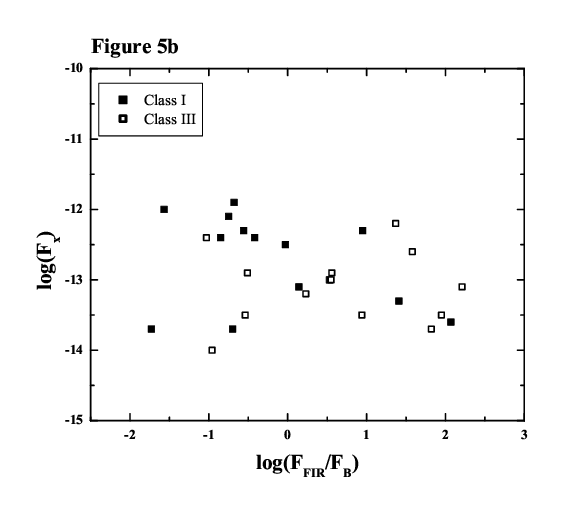}}\\
\caption[]{5a: X-ray flux vs. flux in the far-IR for the expanded sample of LINERs. 5b: X-ray flux vs. IR-brightness for the expanded sample of LINERs.  As can be seen, no correlation is found in either plot.}
\end{figure*}

\subsection{Eddington Ratios and Trends with IR Brightness}

\begin{figure*}[]
\begin{center}
{\includegraphics[width=9.5cm]{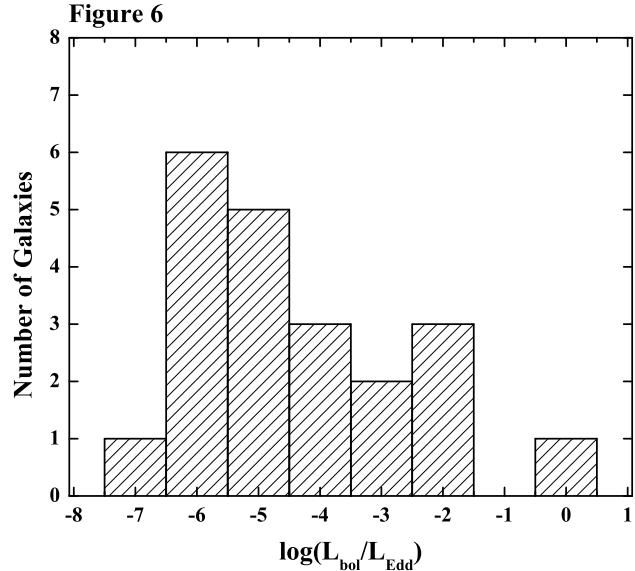}}
\end{center}
\caption[]{Histogram showing the full range of Eddington Ratios for the expanded sample.}
\end{figure*}

Using the X-ray luminosities for our sample of AGN-LINERs, we calculated the corresponding Eddington ratios as outlined in Section 4.  These values are listed in Table 3. In Figure 6, we show the distribution of L/L$_{\rm Edd}$ for the expanded sample.  Consistent with previous studies (e.g. Ho 1999, Terashima et al. 2000), we find that LINERs generally have low Eddington ratios, with a median value of $\sim$ 7 $\times$ 10$^{-6}$ for the expanded sample.  At such low accretion rates ($\leq$ a few percent of the Eddington rate; see Narayan, Mahadevan, \& Quataert 1998) the inner accretion flow is most likely radiatively inefficient (Terashima et al. 2004; Narayan et al. 1998, 2002).

In Figures 7a and 7b, we plot the Eddington ratio as a function of the far-IR luminosity and the IR-brightness ratio, respectively.  Interestingly, we find a surprising trend in L/L$_{\rm Edd}$  vs. both L$_{\rm FIR}$ and L$_{\rm FIR}$/L$_{\rm B}$ that extends over seven orders of magnitude in L/L$_{\rm Edd}$.  A Spearman rank test gives a correlation coefficient of r$_S$=0.64 between L/L$_{\rm Edd}$ and L$_{\rm FIR}$  with a probability of chance correlation of  2$\times$ 10$^{-3}$, indicating a significant correlation between the two values.  In the case of L/L$_{\rm Edd}$ vs. L$_{\rm FIR}$/L$_{\rm B}$, the Spearman rank test gives a correlation coefficient of r$_S$=0.57 with a probability of chance occurrence of 9$\times$10$^{-3}$, also indicating a significant correlation.  

\begin{figure*}[]
{\includegraphics[width=8cm]{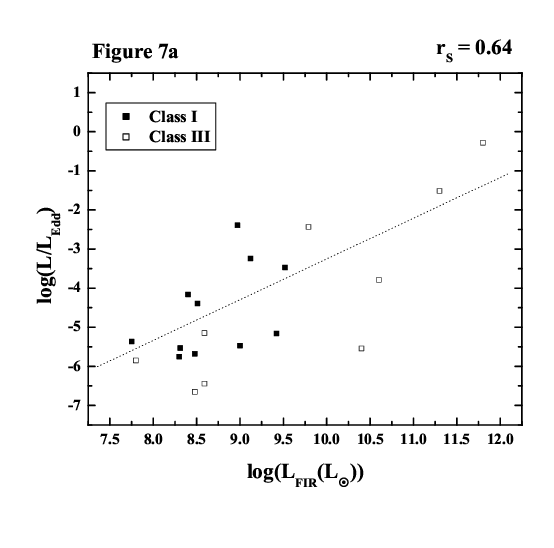}}{\includegraphics[width=8.5cm]{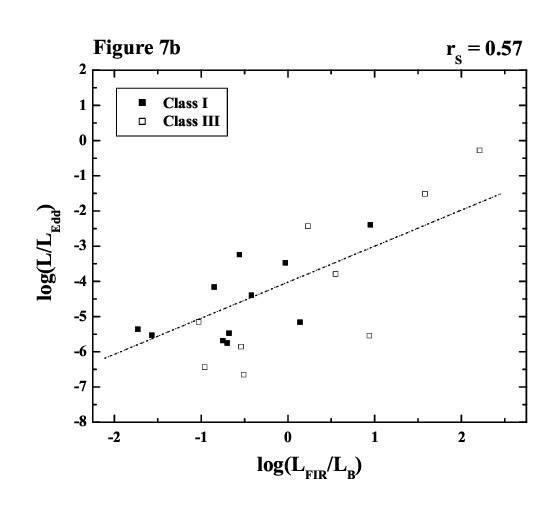}}\\
\caption[]{7a: Eddington Ratio as a function of the luminosity in the far-IR 7b:  Eddington Ratio as a function of IR-brightness.  A significant correlation that extends over seven orders of magnitude in L/L$_{\rm Edd}$ is found in both plots.  The Spearman rank correlation coefficient is given in the upper right corner of each plot.}
\end{figure*}

We investigated whether this correlation is primary or whether it was induced by either distance effects or correlations between individually observed quantities used to calculate the Eddington ratio.  From Section 5.4, we see that both F$_{X}$ vs. F$_{\rm FIR}$ and F$_{\rm X}$  vs. F$_{\rm FIR}$/F$_{\rm B}$ show no correlation, suggesting that Figures 7a and 7b represent a fundamental correlation.  Furthermore, the partial Spearman rank correlation coefficient goes up when the distance is fixed (P$_S$ = 0.66 for L/L$_{\rm Edd}$ vs. L$_{\rm FIR}$ and P$_S$ = 0.67 for L/L$_{\rm Edd}$ vs. L$_{\rm FIR}$/L$_{\rm B}$), suggesting that the Eddington ratio and the IR luminosity and IR-brightness in LINER galaxies are indeed physically correlated quantities.

A formal fit to the correlations in Figures 7a and 7b yield the following relationships:
\begin{equation}
Log(L_{\rm bol}/L_{\rm Edd}) =  (1.04 \pm 0.62)log(L_{\rm FIR}) + (-13.67 \pm 5.40)
\end{equation}
\begin{equation}
Log(L_{\rm bol}/L_{\rm Edd}) = (1.02 \pm 0.38)log(L_{\rm FIR}/L_{B}) + (-4.02 \pm 0.35)
\end{equation}
The dispersion in Figures 7a and 7b is large.  It is difficult to assess how much of the scatter is intrinsic or is due to the uncertainties in the derived quantities.  The bolometric luminosity estimated using the X-ray luminosity is highly uncertain as discussed in Section 4.  In addition, the uncertainty and uniform applicability of the M$_{\rm BH}$ vs. $\sigma$ relationship for our sample of galaxies will introduce additional errors.  In addition, the nonsimultaneity of the X-ray, IR, and optical observations may also introduce additional scatter.  Although the variability properties of low luminosity AGN are not well known, some variability in several sources is found in at least the X-rays (e.g. Ptak et al. 1998).  We regard the correlation in Figures 7a and 7b as preliminary.  A larger sample and more extensive multiwavelength studies of the SEDs of LINERs would provide more accurate bolometric correction factors and allow us to better assess the origin of the scatter in Figure 7.

%----------------------------------------------------------------------
\begin{table*}
\begin{center}
\begin{tabular}{lccccc}
\multicolumn{6}{l}{\bf Table 2: {\it Chandra} Observation Log}  \\ \hline\hline
& & & & &\\
\multicolumn{1}{c}{Galaxy} & OID & Exposure & R. A. & DEC. & Coordinate \\
\multicolumn{1}{c}{Name} & & Time & & & Catalog \\
\multicolumn{1}{c}{(1)} & (2) & (3) & (4) & (5) & (6)\\
& & & & & \\
NGC0660&4010& 5064 &01 43 02.39&+13 38 43.9&{\it 2MASS}\\

NGC1055&4011& 5033 &02 41 45.17&+00 26 38.1&{\it VLA}\\

NGC3125&4012&5153&10 06 33.98&-29 56 17.0&{\it 2MASS}\\

NGC4013&4013&4897&11 58 31.37&+43 56 50.8&{\it 2MASS}\\

NGC4102&4014&4541&12 06 23.05&+52 42 39.7&{\it VLA}\\

NGC4350&4015&4344&12 23 57.82&+16 41 36.1&{\it 2MASS}\\

NGC4419&5283&5061&12 26 56.45&+15 02 50.9&{\it VLA}\\

NGC4527&4017&4897&12 34 08.50&+02 39 13.7&{\it VLA}\\

NGC4666&4018&4642&12 45 8.26&-00 27 50.2&{\it VLA}\\

NGC4713&4019&4904&12 49 57.89&+05 18 41.1&{\it 2MASS}\\

NGC4750&4020&4935&12 50 07.40&+72 52 28.3&{\it 2MASS}\\

NGC5005&4021&4900&13 10 56.28&+37 03 32.4&{\it VLA}\\

NGC5678&4022&4733&14 32 5.84&+57 55 10.0&{\it VLA}\\

NGC5954&4023&4146&15 34 35.16&+15 12 01.5&{\it VLA}\\

IC1218&4024&4638&16 16 37.10&+68 12 09.5&{\it 2MASS}\\

NGC7465&4025&1579&23 02 00.96&+15 57 53.4&{\it 2MASS}\\

\hline
\end{tabular}
\end{center}
{\scriptsize {\bf Column Explanation:} Col(1): Galaxy Common Name; Col(2): {\it Chandra} Observation Identification Number; Col(3): Exposure time in seconds; Col(4):  Right Ascension of nucleus in hours, minutes, \& seconds taken from the source in Column 6;  Col(5): Declination of nucleus in degrees, minutes, \& seconds, taken from the source in Column 6;  Col(6): {\it VLA} Coordinates or {\it 2MASS} coordinates used when extracting counts. {\it VLA} Coordinates came from the First Cataloge search, http://sundog.stsci.edu/cgi-bin/searchfirst.  {\it 2MASS} Coordinates came from NED.}\end{table*}
%----------------------------------------------------------------------

\begin{table*}
\begin{center}
\begin{tabular}{lcccccccc}
\multicolumn{9}{l}{\bf Table 3: X-ray Results of the Expanded Sample}  \\ 
 \hline\hline
\multicolumn{9}{l}{\bf {\it Table 3a: Proprietary X-ray Results}}  \\ \hline
 & & & & & & & & \\
\multicolumn{1}{c}{Galaxy} & X-ray & Hard & Hardness & Count Rate & L$_{\rm X}$ & {\it N}$_{\rm H}$$^{MIN}$ & {\it N}$_{\rm H}$$^{AVG}$ & L$_{\rm bol}$/ \\ 
\multicolumn{1}{c}{Name} & Class & Counts & Ratio & 0.3-10keV & 2-10keV & $\times$10$^{23}$ & $\times$10$^{24}$ & L$_{\rm Edd}$ \\ 
\multicolumn{1}{c}{(1)} & (2) & (3) & (4) & (5) & (6) & (7) & (8) & (9) \\ 
 & & & & & & & & \\
\hline 
NGC0660&II&5&-0.63&0.005 $\pm$0.001&3.3E+38&$\cdots$&$\cdots$&4.0E-06\\

NGC1055&IV&$<$1&$\cdots$&$<$ 0.0004&$<$3.3E+37&6.7&9.7&$\cdots$\\

NGC3125&I&137&-0.25&0.071 $\pm$ 0.004&4.7E+39&$\cdots$&$\cdots$&4.1E-03\\

NGC4013&IV&$<$1&$\cdots$&$<$0.0003&$<$1.6E+37&11.0&10.2&$\cdots$\\

NGC4102&III&80&-0.38&0.057 $\pm$ 0.004&3.1E+39&$\cdots$&$\cdots$&$\cdots$\\

NGC4350&I&10&-0.13&0.005 $\pm$ 0.001&6.4E+38&$\cdots$&$\cdots$&1.8E-06\\

NGC4419&II&20&-0.20&0.010 $\pm$ 0.002&1.2E+39&$\cdots$&$\cdots$&3.9E-05\\

NGC4527&III&14&-0.30&0.008 $\pm$ 0.001&1.9E+39&$\cdots$&$\cdots$&2.9E-06\\

NGC4666&IV&$<$1&$\cdots$&$<$0.0007&$<$1.2E+38&1.1&8.4&$\cdots$\\

NGC4713&IV&$<$1&$\cdots$&$<$0.0004&$<$1.3E+37&12.6&10.4&$\cdots$\\

NGC4750&I&31&-0.58&0.030 $\pm$ 0.003&6.0E+39&$\cdots$&$\cdots$&$\cdots$\\

NGC5005&III&27&-0.76&0.046 $\pm$ 0.003&3.1E+39&$\cdots$&$\cdots$&$\cdots$\\

NGC5678&IV&$<$1&$\cdots$&$<$0.0005&$<$1.5E+38&0.5&8.2&$\cdots$\\

NGC5954&IV&$<$1&$\cdots$&$<$0.0004&$<$1.7E+38&0.3&8.1&$\cdots$\\

IC1218&IV&$<$1&$\cdots$&$<$0.0003&$<$3.5E+37&6.4&9.5&$\cdots$\\

NGC7465$^*$&I&164&$\cdots$&0.160 $\pm$ 0.011&$\cdots$&$\cdots$&$\cdots$&$\cdots$\\
\hline
\end{tabular}
\end{center}
\end{table*}

\begin{table*}
\begin{center}
\begin{tabular}{lcccccccc}
\multicolumn{9}{l}{\bf Table 3: X-ray Results of the Expanded Sample cont.}  \\ \hline\hline
\multicolumn{9}{l}{\bf {\it Table 3b: SSD04 X-ray Statistics}}  \\ \hline
 & & & & & & & & \\
\multicolumn{1}{c}{Galaxy} & X-ray & Hard & Hardness & Count Rate & L$_{\rm X}$ & {\it N}$_{\rm H}$$^{MIN}$ & {\it N}$_{\rm H}$$^{AVG}$ & L$_{\rm bol}$\ \\ 
\multicolumn{1}{c}{Name} & Class$^a$ & Counts$^a$ & Ratio & 0.3-10keV$^a$ & 2-10keV$^a$ & $\times$10$^{23}$ & $\times$10$^{24}$ & L$_{\rm Edd}$ \\ 
\multicolumn{1}{c}{(1)} & (2) & (3) & (4) & (5) & (6) & (7) & (8) & (9) \\ 
 & & & & & & & & \\
\hline 
0248+4302 & II & 1 & -0.85 & 0.001 & 1.8E+40$^a$ & $\cdots$ & $\cdots$ & $\cdots$\\
1346+2650 & III & 52 & -0.81 & 0.027 & 7.1E+41$^a$ & $\cdots$ & $\cdots$ & 1.6E-04\\
2312-5919 & III & 286 & 0.07 & 0.011 & 1.1E+41$^a$ & $\cdots$ & $\cdots$ & $\cdots$\\
2055-4250 & III & 65 & -0.60 & 0.0072 & 7.2E+40$^a$ & $\cdots$ & $\cdots$ & $\cdots$\\
IC1459* & III & 1906 & -0.50 & 0.13 & 2.6E+40$^a$ & $\cdots$ & $\cdots$ & 7.1E-06\\
MRK266 & I & 148 & -0.23 & 0.19 & 7.4E+40$^a$ & $\cdots$ & $\cdots$ & $\cdots$\\
MRK273 & III & 671 & 0.15 & 0.026 & 1.1E+44$^e$ & $\cdots$ & $\cdots$ & 5.3E-01\\
NGC0224* & II & 51 & -0.84 & 0.12 & 3.9E+37$^a$ & $\cdots$ & $\cdots$ & 2.8E-07\\
NGC0253 & II & 254 & -0.13 & 0.042 & 1.1E+38$^a$ & $\cdots$ & $\cdots$ & 2.6E-06\\
NGC0404 & II & 25 & -0.72 & 0.0076 & 2.1E+37$^a$ & $\cdots$ & $\cdots$ & 7.5E-06\\
NGC0835 & II & 15 & -0.47 & 0.0045 & 7E+39$^a$ & $\cdots$ & $\cdots$ & 2.0E-06\\
NGC1052* & I & 183 & 0.24 & 0.13 & 4.2E+41$^b$ & $\cdots$ & $\cdots$ & 5.8E-04\\
NGC3031 & I & 76 & -0.32 & 0.093 & 1.6E+40$^c$ & $\cdots$ & $\cdots$ & 6.9E-05\\
NGC3079 & II & 95 & -0.08 & 0.0078 & 6.8E+39$^a$ & $\cdots$ & $\cdots$ & 4.1E-06\\
NGC3368 & II & 1 & -0.78 & 0.0045 & 2.8E+39$^a$ & $\cdots$ & $\cdots$ & 5.3E-05\\
NGC3623 & IV & $<$2 & $\cdots$ & $<$0.0081 & $<$4E+38$^a$ & N/A & 7.2 & $\cdots$\\
NGC4125 & III & 35 & -0.74 & 0.0042 & 4.2E+38$^a$ & $\cdots$ & $\cdots$ & 3.6E-07\\
NGC4278* & I & 35 & -0.68 & 0.21 & 1.2E+40$^c$ & $\cdots$ & $\cdots$ & 2.1E-06\\
NGC4314 & II & 5 & -0.70 & 0.0021 & 1.4E+38$^a$ & $\cdots$ & $\cdots$ & 2.3E-06\\
NGC4374 & III & 153 & -0.70 & 0.036 & 1.3E+39$^c$ & $\cdots$ & $\cdots$ & 2.2E-07\\
NGC4486* & I & 1948 & -0.66 & 0.3 & 3.3E+40$^a$ & $\cdots$ & $\cdots$ & 2.9E-06\\
NGC4569 & II & 10 & -0.60 & 0.0296 & 2.6E+39$^c$ & $\cdots$ & $\cdots$ & 1.8E-05\\
NGC4579* & I & 8278 & -0.39 & 0.81 & 8.9E+40$^c$ & $\cdots$ & $\cdots$ & 3.4E-04\\
NGC4696 & III & 39 & -0.59 & 0.0022 & 1.3E+40$^a$ & $\cdots$ & $\cdots$ & 8.9E-06\\
NGC5194 & III & 49 & -0.79 & 0.018 & 1.1E+41$^d$ & $\cdots$ & $\cdots$ & 3.7E-03\\
NGC5195 & IV & $<$1 & $\cdots$ & $<$0.0009 & $<$7.1E+37$^c$ & 3.4 & 8.9 & $\cdots$\\
NGC6240 & III & 1110 & 0.04 & 0.058 & 1.6E+44$^f$ & $\cdots$ & $\cdots$ & 3.1E-02\\
NGC6500 & I & 1 & -0.95 & 0.02 & 1.7E+40$^a$ & $\cdots$ & $\cdots$ & 6.96E-06\\
NGC6503 & II & 11 & -0.42 & 0.0029 & 4.6E+35$^a$ & $\cdots$ & $\cdots$ & 3.7E-07\\
NGC7331 & II & 33 & -0.56 & 0.0051 & 3.3E+38$^a$ & $\cdots$ & $\cdots$ & 1.1E-06\\
UGC05101 & I & 150 & -0.15 & 0.0072 & 7.7E+40$^a$ & $\cdots$ & $\cdots$ & $\cdots$\\

\hline
\end{tabular}
\end{center}
\end{table*}

\begin{table*}
\begin{center}
\begin{tabular}{lcccccccc}
\multicolumn{9}{l}{\bf Table 3: X-ray Results of the Expanded Sample cont.}  \\\hline\hline
\multicolumn{9}{l}{\bf {\it Table 3c: H01 X-ray Statistics}}  \\ \hline
 & & & & & & & & \\
\multicolumn{1}{c}{Galaxy} & X-ray & Hard & Hardness & Count Rate & L$_{\rm X}$ & {\it N}$_{\rm H}$$^{MIN}$ & {\it N}$_{\rm H}$$^{AVG}$ & L$_{\rm bol}$\ \\ 
\multicolumn{1}{c}{Name} & Class & Counts & Ratio & 0.3-10keV & 2-10keV & $\times$10$^{23}$ & $\times$10$^{24}$ & L$_{\rm Edd}$ \\ 
\multicolumn{1}{c}{(1)} & (2) & (3) & (4) & (5) & (6) & (7) & (8) & (9) \\ 
 & & & & & & & & \\
\hline 
NGC 2787 & III & NA & NA & 0.0053 & 2.0E+38 & $\cdots$ & $\cdots$ & 1.4E-06\\
NGC 2841 & II & NA & NA & 0.0035 & 1.8E+38 & $\cdots$ & $\cdots$ & 1.9E-07\\
NGC 3489 & II & NA & NA & 0.0061 & 1.7E+38 & $\cdots$ & $\cdots$ & 1.2E-06\\
NGC 3627 & IV & NA & NA & $<$0.0017 & $<$4.1E+37 & 5.5 & 9.3 & $\cdots$\\
NGC 3628 & IV & NA & NA & $<$0.0017 & $<$4.8E+37 & 4.7 & 9.2 & $\cdots$\\
NGC 3675 & IV & NA & NA & $<$0.0017 & $<$9.8E+37 & 1.6 & 8.5 & $\cdots$\\
NGC 4203 & I & NA & NA & 0.22 & 1.2E+40 & $\cdots$ & $\cdots$ & 4.1E-05\\
NGC 4321 & IV & NA & NA & $<$0.0040 & $<$3.9E+38 & NA & 7.3 & 1.6E-05\\
NGC 4494 & I & NA & NA & 0.012 & 7.2E+38 & $\cdots$ & $\cdots$ & 4.4E-06\\
NGC 4594 & I & NA & NA & 0.19 & 1.4E+40 & $\cdots$ & $\cdots$ & 3.4E-06\\
NGC 4826 & IV & NA & NA & $<$0.0067 & $<$7.2E+37 & 2.8 & 8.8 & $\cdots$\\
\hline
\end{tabular}
\end{center}

{\noindent{\scriptsize {\bf Column Explanation:} Col(1): Galaxy Common Name, ($^*$) galaxies experiencing pileup; Col(2): X-ray morphological Class; Col(3): Hard counts in the nucleus (2-8 keV) from an extraction region of radius 2'' centered on the radio or {\it 2MASS} nucleus, NA=Not Available; Col(4): Hardness Ratio:  Defined here as (H-S)/(H+S) where H represents the hard counts (2-8 kev) and S represents the soft counts (0.3-2 keV) in the nucleus, NA=Not Available.  Col(5): Countrate (counts/sec), NGC 7465 experienced severe pileup and could not be confidently analyzed; Col(6): X-ray luminosity in ergs s$^{-1}$; Col(7):Intrinsic absorption corresponding to upper limits for all non-detections in units of cm$^{-2}$.  This is the column density required to obscure an AGN with luminosity equal to L$_{\rm X}$ = 2 $\times$ 10$^{38}$ ergs s$^{-1}$, which is the minimum 2-10keV luminosity of all targets with detections of hard nuclear sources from this sample, H01, and SSD04; Col(8):Intrinsic absorption corresponding to upper limits for all non-detections in units of cm$^{-2}$.  This is the column density required to obscure an AGN with luminosity equal to L$_{\rm X}$ = 9.5 $\times$ 10$^{42}$ ergs s$^{-1}$, which is the average 2-10keV luminosity of all targets with detections of hard nuclear sources from this sample, H01, and SSD04; Col(9): Eddington ratio}}
{\noindent{\scriptsize {\bf References:} $^a$ SSD04, $^b$ Guainazzi et al 2000, $^c$ Ho et al. 2001, $^d$ Fukazawa et al. 2001, $^e$ Xia et al. 2001, $^f$ Vignati et al. 1999.}} 

\end{table*}

\section{Discussion}
\subsection{Possible Connection Between Black Hole Growth and Star Formation?}

Figures 7a and 7b imply that either the mass accretion rate, or the radiative efficiency, or a combination of both, scales with the IR luminosity and IR-brightness of the galaxy.  The majority of galaxies plotted in Figure 7 have extremely low accretion rates, well below the threshold at which the accretion flow is likely to take place via an Advection Dominated Accretion Flow (ADAF) model or other radiatively inefficient accretion models (Narayan, Mahadevan, \& Quataert 1998).  The smooth increase in L/L$_{\rm Edd}$ with respect to L$_{\rm FIR}$/L$_{\rm B}$ argues against the possibility that the accretion mode and therefore the radiative efficiency, changes significantly for the targets plotted in Figure 7. For operative purposes, in this paper, we assume constant radiative efficiency and explore the consequences.

The IR luminosities, as measured from IRAS, as well as the blue magnitudes are measured in a large aperture and therefore include the emission from the entire host galaxy in our sample of LINERs.  The infrared emission from galaxies is usually attributed to thermal radiation from dust heated by young massive stars, the underlying old stellar population, and possibly an AGN (Rowan-Robinson \& Crawford 1989).  In the absence of an AGN, both the far-IR luminosity as well as the L$_{\rm FIR}$/L$_{\rm B}$ ratio are widely used as star formation indicators (e.g., Keel 1993, Huang et al. 1996, Hunt \& Malkan 1999).  In starburst galaxies, both quantities can be used as a direct measure of the star formation rate (SFR; Lehnert \& Heckman 1996; Meurer et al. 1997).  Even in the case of early-type spiral galaxies, a comparison of H$\alpha$ equivalent widths with IR-brightness suggests that the far-IR luminosity and L$_{\rm FIR}$/L$_{\rm B}$ ratio can be used as a reliable star formation indicator (Usui, Saito, \& Tomita 1998).  In our sample of galaxies, the contributions to the IR emission from underlying old stars, the AGN, and the fraction of stellar light reprocessed by dust is unknown.   However, given the low X-ray luminosities and Eddington ratios for our sample of LINERs, the contribution from the AGN to the far-IR luminosity is likely to be small.  If we make the assumption that far-IR emission is predominantly associated with young massive stars, Figure 7 implies a link between black hole growth, as measured by the mass accretion rate, and the SFR, as measured by the far-IR luminosity and IR-brightness.

	A correlation between mass accretion rate and SFR in LINERs may have important implications for our understanding of black hole growth and the connection between starbursts and AGN.  The well-known tight correlation between black hole mass and stellar velocity dispersion (Gebhart et al. 2000) implies that black hole and bulge formation and growth are intimately connected.  In a study of narrow line Seyfert 1 (NLS1s) galaxies, Shemmer et al. (2004) find that the Eddington ratio is correlated with gas metallicity, implying a similar relationship between star formation, responsible for the gas metal enrichment, and the accretion rate.  Unlike NLS1s, thought to represent an early phase in the lifetime of an AGN, LINERs are characterized by more massive black holes and low accretion rates, possibly representing the phase in galaxy evolution just before accretion ``turns off".  Figure 7 may suggest an evolutionary sequence, where IR-bright LINERs represent younger AGN defined by higher mass accretion rates, evolving into lower accretion rate IR-faint LINERs, and finally, the normal quiescent galaxies with dormant black holes we see in our local Universe.  Testing for systematic differences in the circumnuclear stellar populations in LINERs may prove vital to our understanding of galaxy evolution.

\subsection{Speculations on AGN Fueling}

The far-IR luminosity or H$\alpha$ equivalent width is known to be roughly proportional to the CO emission from galactic and extragalactic molecular clouds (e.g., Mooney \& Solomon 1988 ; Devereux \& Young 1991; Young et al. 1996; Rownd \& Young 1999) suggesting that the SFR rate in galaxies is proportional to the molecular gas mass (e.g., Mooney \& Solomon 1988).  CO-line observations for the majority of our sample do not exist in the literature.  If the scaling between the far-IR luminosity and the CO emission holds for the host galaxies in our sample, Figure 7a suggests the intriguing possibility that the mass accretion rate scales with the host galaxy's fuel supply.

AGN fueling has been a longstanding question.  While it is widely accepted that the onset of activity results from the feeding of the black hole by the gas reservoir of the host galaxy, there is no consensus on which mechanism is responsible for removing angular momentum and driving the gas infall down to scales of less than a parsec (see reviews by Martini 2004 and Wada 2004).  A number of mechanisms for removing the angular momentum of the gas have been proposed.  Among these mechanisms, galaxy interactions (eg Toomre \& Toomre, 1972; Shlosman et al. 1989, 1990; Hernquist \& Mihos, 1995) and galactic bars (e.g. Schwarz 1984; Shlosman, Frank \& Begelman 1989; Knapen et al. 1995) are prime candidates for facilitating the transfer of mass from large to small scales.   However, despite many years of observational effort, no strong correlations between the presence of any of the proposed large scale fueling mechanisms in low luminosity AGN have been found.  For example, the majority of recent observations show either a marginal or no clear excess of bars or interactions in galaxies harboring low luminosity AGN as compared with normal galaxies (e.g. Ho, Fillipenko, \& Sargent 1997a, Mulchaey \& Regan 1997, Corbin 2000, Schmitt  2001).  If Figure 7 implies that the mass accretion rate scales with fuel supply over seven orders of magnitude in L/L$_{Edd}$, then the fueling mechanism responsible must be compatible with this result.  Of the targets plotted in Figure 7, two of the LINERs contain bars and the two with the most extreme L$_{\rm FIR}$ and L$_{\rm FIR}$/L$_{\rm B}$ values are interacting systems, suggesting that the large scale phenomenon does not have a significant impact on the fueling of the AGN .  If different mechanisms are responsible for AGN fueling over the range of Eddington ratios plotted in Figure 7, it would be unexpected to see such a smooth trend in mass accretion rate with fuel supply.  Rather one would expect the galaxies in Figure 7 to occupy well-defined regimes in L/L$_{\rm Edd}$ vs. L$_{\rm FIR}$ or L$_{\rm FIR}$/L$_{\rm B}$ based on the fueling mechanism responsible for accretion onto a black hole.  We reiterate that the conclusions drawn from Figure 7 are highly speculative and based on limited data.  More rigorous and extensive multi-wavelength observations are critical to validate our results.

\section{Summary and Conclusions}
 
We have examined the properties of a sample of nearby IR-bright LINERs using proprietary and archival X-ray observations from {\it Chandra}.  Our main results are as follows. 

\begin{enumerate}
\item Twenty-eight out of 55 LINERs (51\%) show compact hard nuclear cores coincident with the radio or {\it 2MASS} nucleus, with a luminosity L$_X$ (2-10keV) $\geq$ 2 $\times$ 10$^{38}$ ergs s$^{-1}$.  The nuclear 2-10 keV luminosity for this expanded sample of galaxies ranges from $\sim$ 2 $\times$ 10$^{38}$ ergs s$^{-1}$ to $\sim$ 2 $\times$ 10$^{44}$ ergs s$^{-1}$.   \item We find that the most IR-faint LINERs are exclusively AGN.  The fraction of LINERs containing AGN appears to decrease with IR-brightness and increase again at the highest values of L$_{\rm FIR}$/L$_{\rm B}$.  \item Of the LINERs with hard X-ray nuclear cores and observations at H$\alpha$ wavelengths, only 33\% (8/24) show broad lines.  Similarly, only roughly half of the LINERs observed in the radio (8/14) show a flat spectrum radio core.  These findings emphasize the need for high-resolution X-ray imaging observations in the study of IR bright LINERs.   \item We find a surprising trend in the Eddington ratio as a function of L$_{\rm FIR}$/L$_{\rm B}$ and L$_{\rm FIR}$ that extends over seven orders of magnitude in L/L$_{\rm Edd}$. This may imply a fundamental link between mass accretion rate, as measured by the Eddington ratio, and star formation rate (SFR), as measured by the IR-brightness.  The correlation may further indicate that the accretion rate scales with fuel supply, a finding that has important implications for AGN fueling in low luminosity AGN.  
\end{enumerate}

We are very thankful to Davide Donato, Alex Rinn \& Brian O'Halloran for their invaluable help in the data analysis, and for numerous instructive discussions and to Joe Weingartner, for his enlightening and thoughtful comments.  We are also very grateful for the insightful suggestions from the referee, which helped improve this work.  This research has made use of the NASA/IPAC Extragalactic Database (NED) which is operated by the Jet Propulsion Laboratory, California Institute of Technology, under contract with the National Aeronautics and Space Administration.  SS gratefully acknowledges financial support from NASA grant NAG5-11432 and NAG03-4134X.  RMS gratefully acknowledges financial support from NASA LTSA grant NAG5-10708 and from the Clare Boothe Luce Program of the Henry Luce Foundation.

\end{document}